%% file: main.tex
\documentclass[journal= revtex4,manuscript=article]{achemso}
\mciteErrorOnUnknownfalse
\setkeys{acs}{articletitle = true} 
\usepackage{graphicx}
\usepackage{dcolumn}
\usepackage{bm}
\graphicspath{ {figures/} }
\usepackage{hyperref}
\usepackage{chemformula}
\usepackage{notes2bib}
\usepackage{times}

\newcommand{\unit}[1]{\ensuremath{\, \mathrm{#1}}}

\title{Room Temperature Fiber-Coupled single-photon devices based on Colloidal Quantum Dots and SiV centers in Back Excited Nanoantennas}

\author{Boaz Lubotzky}
\affiliation{Racah Institute of Physics, The Hebrew University of Jerusalem, Jerusalem 9190401, Israel}
\affiliation{The Center for Nanoscience and Nanotechnology, The Hebrew University of Jerusalem, Jerusalem 9190401, Israel}
\author{Alexander Nazarov}
\affiliation{Racah Institute of Physics, The Hebrew University of Jerusalem, Jerusalem 9190401, Israel}
\affiliation{The Center for Nanoscience and Nanotechnology, The Hebrew University of Jerusalem, Jerusalem 9190401, Israel}
\author{Hamza Abudayyeh}
\affiliation{Racah Institute of Physics, The Hebrew University of Jerusalem, Jerusalem 9190401, Israel}
\affiliation{The Center for Nanoscience and Nanotechnology, The Hebrew University of Jerusalem, Jerusalem 9190401, Israel}
\author{Lukas Antoniuk}
\affiliation{Institute for Quantum Optics, University of Ulm, Albert-Einstein-Allee 11,89081 Ulm, Germany}
\author{Niklas Lettner }
\affiliation{Institute for Quantum Optics, University of Ulm, Albert-Einstein-Allee 11,89081 Ulm, Germany}
\affiliation{Center for Integrated Quantum Science and Technology (IQst), Ulm University, Albert-Einstein-Allee 11, D-89081 Ulm, Germany}
\author{Viatcheslav Agafonov }
\affiliation{GREMAN, UMR 7347 CNRS, INSA-CVL, Tours University, 37200 Tours,France}
\author{Anastasia V. Bennett}
\affiliation{Materials Physics \& Applications Division: Center for Integrated Nanotechnologies, Los Alamos National Laboratory,
Los Alamos, New Mexico 87545, USA}\author{Somak Majumder}
\affiliation{Materials Physics \& Applications Division: Center for Integrated Nanotechnologies, Los Alamos National Laboratory,
Los Alamos, New Mexico 87545, USA}\author{Vigneshwaran Chandrasekaran}
\affiliation{Materials Physics \& Applications Division: Center for Integrated Nanotechnologies, Los Alamos National Laboratory,
Los Alamos, New Mexico 87545, USA}\author{ Eric G. Bowes}
\affiliation{Materials Physics \& Applications Division: Center for Integrated Nanotechnologies, Los Alamos National Laboratory,
Los Alamos, New Mexico 87545, USA}\author{Han Htoon}
\affiliation{Materials Physics \& Applications Division: Center for Integrated Nanotechnologies, Los Alamos National Laboratory,
Los Alamos, New Mexico 87545, USA}
\author{Jennifer A. Hollingsworth}
\affiliation{Materials Physics \& Applications Division: Center for Integrated Nanotechnologies, Los Alamos National Laboratory,
Los Alamos, New Mexico 87545, USA}
\author{Alexander Kubanek}
\affiliation{Institute for Quantum Optics, University of Ulm, Albert-Einstein-Allee 11,89081 Ulm, Germany}
\affiliation{Center for Integrated Quantum Science and Technology (IQst), Ulm University, Albert-Einstein-Allee 11, D-89081 Ulm, Germany}
\author{Ronen Rapaport}
\affiliation{Racah Institute of Physics, The Hebrew University of Jerusalem, Jerusalem 9190401, Israel}
\affiliation{The Center for Nanoscience and Nanotechnology, The Hebrew University of Jerusalem, Jerusalem 9190401, Israel}
\email{ronenr@phys.huji.ac.il}

\begin{document}

\maketitle
\graphicspath{ {Figures/}} 
\begin{abstract}
We demonstrate an important step towards on-chip integration of single-photon sources operating at room temperature—fiber coupling of a directional quantum emitter with back-excitation. Directionality is achieved with a hybrid metal-dielectric bullseye antenna, while back-excitation is permitted by placement of the emitter at or in a sub-wavelength hole positioned at the bullseye center. Overall, the unique design enables a direct laser excitation from the back of the on-chip device and very efficient coupling of the highly collimated photon emission to either low numerical aperture (NA) free-space optics or directly to an optical fiber from the front. To show the versatility of the concept, we fabricate devices containing either a colloidal quantum dot or a silicon-vacancy center containing nanodiamond, which are accurately coupled to the nano-antenna using two different nano-positioning methods. Both back-excited devices display front collection efficiencies of $\sim$70$\%$ at NAs as low as 0.5. Moreover, the combination of back-excitation with forward low-NA directionality enables direct coupling of the emitted photons into a proximal optical fiber without the need of any coupling optics, thereby facilitating and greatly simplifying future integration.

\end{abstract}

\section{Introduction}

Solid-state based single-photon sources (SPS) are the key for a host of quantum technologies. They serve as essential building blocks for quantum metrology\cite{Giovannetti2004Quantum-enhancedLimit,VonHelversen2019QuantumDetectors,Fiderer2018QuantumSensors}, quantum networks\cite{Aharonovich2016Solid-stateEmitters, Nemoto2016PhotonicCenters,Ruf2021QuantumDiamond,Zhang2022On-chipCircuits} and quantum information processing\cite{Kok2007LinearQubits,Arakawa2020ProgressOverview,Gould2016Large-scaleInformation}. 
For room temperature (RT) operation of SPS, colloidal quantum dots (CQDs) are now capable of single-photon emission on-demand  with high brightness, good photon purity,  and a tunable spectrum. Color centers in diamonds display a very stable single-photon emission\cite{Jelezko2006SingleReview,Aharonovich2014,Aharonovich2016Solid-stateEmitters}. Specifically the negatively-charged silicon vacancy center in diamond (SiV) combines a short lifetime, and a narrow band, linearly polarized emission\cite{Lagomarsino2021CreationImplantation}. 
However, a major problem of these sources is their isotropic emission which results in low collection efficiency  (CE), severely limiting their usefulness. 
Modifying the photonic environment near the emitter using dielectric or metallic antenna structures can change the angular emission pattern \cite{Pelton2015ModifiedStructures,Dey2016PlasmonicDots} and lead to higher collection efficiencies. 
Coupling the emitter to high quality-factor dielectric antennas enables low loss and high CE, but is suitable only for narrow frequency bandwidth which, typically, does not cover the bandwidth of RT emitters and requires complex fabrication. Another possibility is to couple to metallic antenna structures. The combination of low-quality factors with low mode volumes enables high directional emission over a broad spectral range. However, efficient coupling requires a short distance between the emitter and the metal, which produces high non-radiative recombination rates\cite{Dey2016PlasmonicDots,Andreussi2004RadiativeShape}. 
A possible solution that combines the advantages of a dielectric and metallic antennas, such as high directivity over broad frequency bandwidth with low loss is the use of a hybrid metal-dielectric circular Bragg (bullseye) antenna designs \cite{Abudayyeh2017}. This concept has been utilized in recent years to demonstrate highly directional RT single-photon emission from a single CQD\cite{Livneh2016HighlyDevice}, and from a single NV center in a nanodiamond \cite{Nikolay2018AccurateStructuresb}, with record-high, near unity collection efficiencies \cite{Abudayyeh2021SingleNanoantennas} combined with a large optical Purcell factor to yield both a high brightness and excellent directionality of single-photons \cite{Abudayyeh2021OvercomingSource}. 

However, one major challenge on the route to a compact, on-chip RT SPS device that can be scaled up and integrated into real world quantum optical applications, is the need for a direct coupling of the emission into an optical fiber. In addition, a pathway to integrating the laser excitation source of the emitter compactly into the same SPS device is needed for full on-chip integration. 
Recently, several works presented a fiber-coupled single-photon source, but only in a cryogenic environment
\cite{Schlehahn2018,Musia2020PlugPlayO-Band,Stein2020Narrow-BandSource,Daveau2016b,Knall2022EfficientSystem,XU2022BrightResonator,Jeon2022Plug-and-PlayInterface}.
In Ref. \cite{Schlehahn2018} for example, the excitation of the emitter and the emission collection is achieved through the same optical fiber. The cryogenic temperature makes it possible to obtain a narrow emission spectrum of the quantum emitter (an epitaxial QD in this case) and by spectral filtering to obtain single-photons with low noise.  This is still a challenge for a RT SPS device however, as the broadband auto-fluorescence of the fiber \cite{} arising from the strong excitation laser light overwhelms the single-photon emission, which is hard to separate due to the more broadband nature of RT SPS.

In this work we address this challenge by avoiding the need for the single-photons and the excitation laser to share the same optical path or the same fiber. Our versatile SPS device consists of a RT quantum emitter coupled to a metal-dielectric hybrid bullseye nano-antenna having a sub-wavelength hole at its center. This geometry enables a direct laser excitation from the back of the on-chip device and direct coupling of the photon emission to either free space optics or directly to an optical fiber from the front. It also results in highly directional forward propagating photons, with negligible back-leak.
To show the versatility of our concept, we demonstrate fabrication of nanoantenna SPS devices containing two different photon sources, namely CQD and SiV in nanodiamonds, which are accurately coupled to the nanoantenna using two different nano-positioning methods, dip-pen nanolithography (DPN) and pick-and-place techniques, respectively.
 Significantly, both devices produce a highly directional forward photon emission that results in CE of $\sim$70$\%$ at numerical apertures (NA) as low as 0.5, while using direct excitation from the back. Moreover, the combination of back-excitation with forward low NA directionality enables a coupling of the emitted photons directly into a proximal optical fiber without the need of further coupling optics, facilitating and greatly simplifying future integration into envisioned photonic quantum networks. 

\section{RESULTS AND DISCUSSION}
\label{:results}
\label{Results}
\subsection{Nano-antenna design and fabrication}

\begin{figure}[H]
\centering
 \includegraphics[width=85mm,scale=0.75]{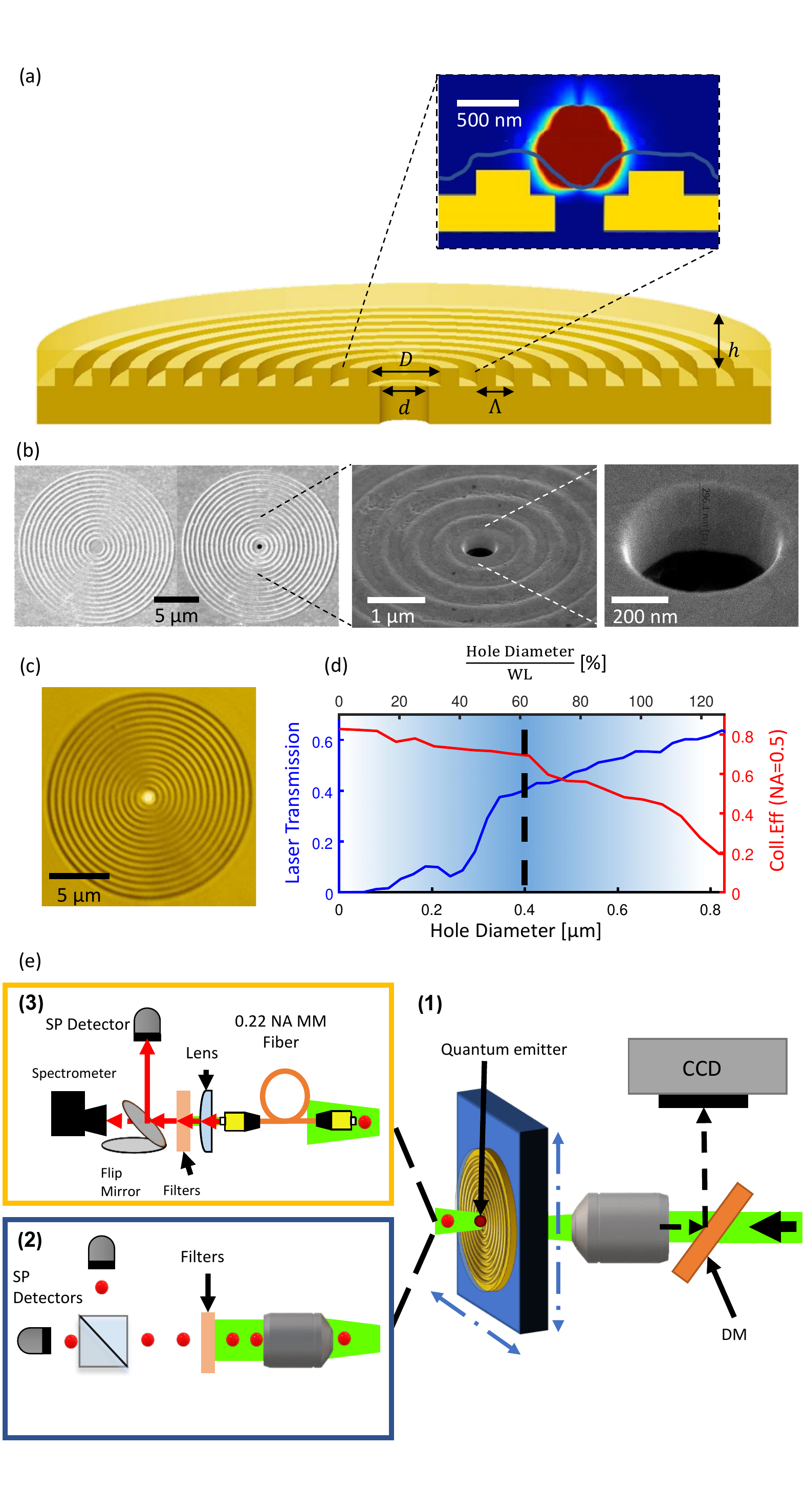}

 \caption{Device schematics and experimental setups.(a) Schematic cross section of hybrid metal-dielectric bullseye structure with a hole at the center with the key geometric parameters: the central cavity diameter (D), hole diameter(d), groove period ($\Lambda$), polymer thickness (h) and total metal thickness (l).
  The inset illustrates a zoom in of the electric field distribution 3D Lumerical simulation of a dipole source located in the center of the antenna, above the hole. The blue line marks the boundary of the layer of aluminium oxide as measured on a sample using AFM. It can be seen that a hole with a diameter of 400 nm does not allow the transmission of the light emitted from the dipole source. (b) SEM images of the bullseye antenna before and after digging the hole. 
(c) Microscope image of the bullseye antenna showing light coming from the hole at the center. 
(d) Laser transmission through the hole and collection efficiency at different hole diameters. The top axis displays the diameter of the hole divided by the wavelength of the emission. The dotted black line marks the actual diameter we used for this study. 
(e) Illustration of the experimental setup: (1) The back excitation system with a dichroic mirror (DM) on the laser path for back imaging, (2) front imaging system, (3) fiber coupling system.    
}

\label{fig:A}
\end{figure}

An illustration of a cross section of the nanoantenna device is seen in Fig. \ref{fig:A}a. The design concept \cite{Livneh2016HighlyDevice,Abudayyeh2017,Abudayyeh2021SingleNanoantennas} and fabrication method are similar to our previous hybrid metal-dielectric nano-antenna devices \cite{Abudayyeh2021SingleNanoantennas}. It consists of a gold substrate (with a thickness of $H$) with a gold circular Bragg grating of height $l$ and period $\Lambda$ surrounding a central circle with a diameter $D$. The metal part is covered by a transparent dielectric layer of aluminum oxide with a height $h$. 
The geometrical parameters of the antenna were numerically optimized \cite{Abudayyeh2017,Abudayyeh2021SingleNanoantennas} (using 3D FDTD Lumerical simulations) for best directionality of the emission from a dipole emitter placed inside the dielectric layer at the center of the antenna. A different set of parameters are found for different emitter central wavelengths, corresponding to the two different emitters used in this work, namely CQDs (650 nm) and SiV (737 nm). The optimized values are given in Appendix A. 
The template stripping method used to fabricate the metalic structure ensures high quality bullseye antennas,
as reported previously \cite{Abudayyeh2021OvercomingSource,Abudayyeh2021SingleNanoantennas}.

Unlike our previous designs, before covering the metal with a dielectric layer, a hole with a diameter $d=400$ nm is drilled at the center using focused ion beam (FIB), penetrating all the way to the transparent substrate. 
See Appendix A for details. 
  
Fig. \ref{fig:A}b shows the bullseye antenna before and after the hole drilling.
In Fig. \ref{fig:A}c, a microscope image of the device with white light passing through the hole in the center of the antenna is shown.

The hole size must be large enough to allow transmission of the laser light from the back of the device through the hole, \cite{Bethe1944TheoryHoles,MorrisJ.Ehrlich2004StudiesField, Guha2005DescriptionTheory} and small enough to maintain the CE by reducing the transmission of the light emitted from the dipole emitter (as illustrated in the inset of Fig. \ref{fig:A}a). This allows coupling to the forward propagating modes of the antenna.
To determine the optimal hole size for our device, we performed calculations based on FDTD simulations. 
Fig.\ref{fig:A}d shows the calculated dependence of the laser transmission through the hole and the far field CE of the dipole emitter  (for NA = 0.5) on the diameter of the hole.
Here, we show the laser transmission for a wavelength of 532 nm, which is the longest wavelength we used for excitation. For a shorter wavelength even more laser excitation power can pass through the hole. As for the CE, we show it for an emitter at  650 nm which is the shorter of the two central wavelengths corresponding to the two nano-emitter types measured in the experiment. Here again, for a longer wavelength, even less emission is expected to leak back through the hole (see Ref.  \cite{Bethe1944TheoryHoles,MorrisJ.Ehrlich2004StudiesField, Guha2005DescriptionTheory}).
 The dotted black line marks the size of the hole (d = 400nm) that was chosen, for which more than 40$\%$ of the back excitation  is transmitted and the CE for NA = 0.5 is larger than 70$\%$. 
Before placing the emitter on the antenna the sample was covered with a dielectric layer of \ch{Al2O3} using ALD deposition (see the blue line in the inset of Fig. \ref{fig:A}a that indicates the boundary of the layer of aluminum oxide as measured on a sample using AFM). The thickness of the oxide layer was chosen to optimize collection efficiency for each emitter type. In this way, the device is easily tailored to the particular single-photon source. 
\subsection{Optical setup}

Typically\cite{Harats2014,Livneh2016HighlyDevice, Abudayyeh2021OvercomingSource,Abudayyeh2021SingleNanoantennas,Nikolay2018AccurateStructures,Waltrich2021High-purityAntenna}, measurements are done in a front excitation configuration, where the optical path of the laser excitation coincides with that of the emission collection. In that case, the same objective is used for both excitation and collection of photons from the quantum emitter. 
 In contrast, in this work a back-excitation/front-collection scheme is enabled by the unique antenna design. As presented schematically in Fig. \ref{fig:A}e the laser light is focused onto the sample using a back  objective (1) to excite the quantum emitter through the hole in the nano-antenna. A CCD camera images the back side of the antenna to align the excitation path. From the front side of the sample, a collection objective (2) or a fiber (3) is used to collect the emission of the quantum emitter. After the light passes through the objective or the fiber, dichroic filters are used to filter out the excitation light and the photons emitted from the emitter reach the detection arms, consisting of either a spectrometer with a CCD camera or a single-photon detector setup capable of measuring emission lifetime traces or second order photon correlations.

Importantly, the two distinct optical paths allow the excitation conditions (e.g., position, intensity, incoming polarization, and spot size) to be varied separately without changing the collection path.
For example, an objective with low NA can be used to enlarge the excitation spot size on the back of the sample, relaxing exact positioning of the excitation spot, and allowing several nano-antennas to be excited at the same time without sacrificing the NA of the collection objective. In addition, the optical components can be separately chosen to be optimal to the specific wavelength and polarization of the laser used for excitation without restrictions imposed by the optimization of the emission collection. The flexibility of such a dual optical system is thus suitable for a greater variety of emitters and conditions, as is shown below.
Beyond the above advantages, in this geometry optical fibers can be used to directly collect the front photon emission, without the requirement that the same fibers should also be used to excite the emitter. This advantage is important as direct excitation through the fiber requires particularly high intensities due to its low NA compared to high NA objectives, which in turn adds to the noise in the collection of the emitted photons due to the strong broadband auto-fluorescence of the fiber.


 

\subsection{Fabrication and characterization of a single-photon device based on giant CQDs}
\begin{figure}[H]
\centering
\includegraphics[width=.7\textwidth,trim={0 2.5cm 0 0},clip]{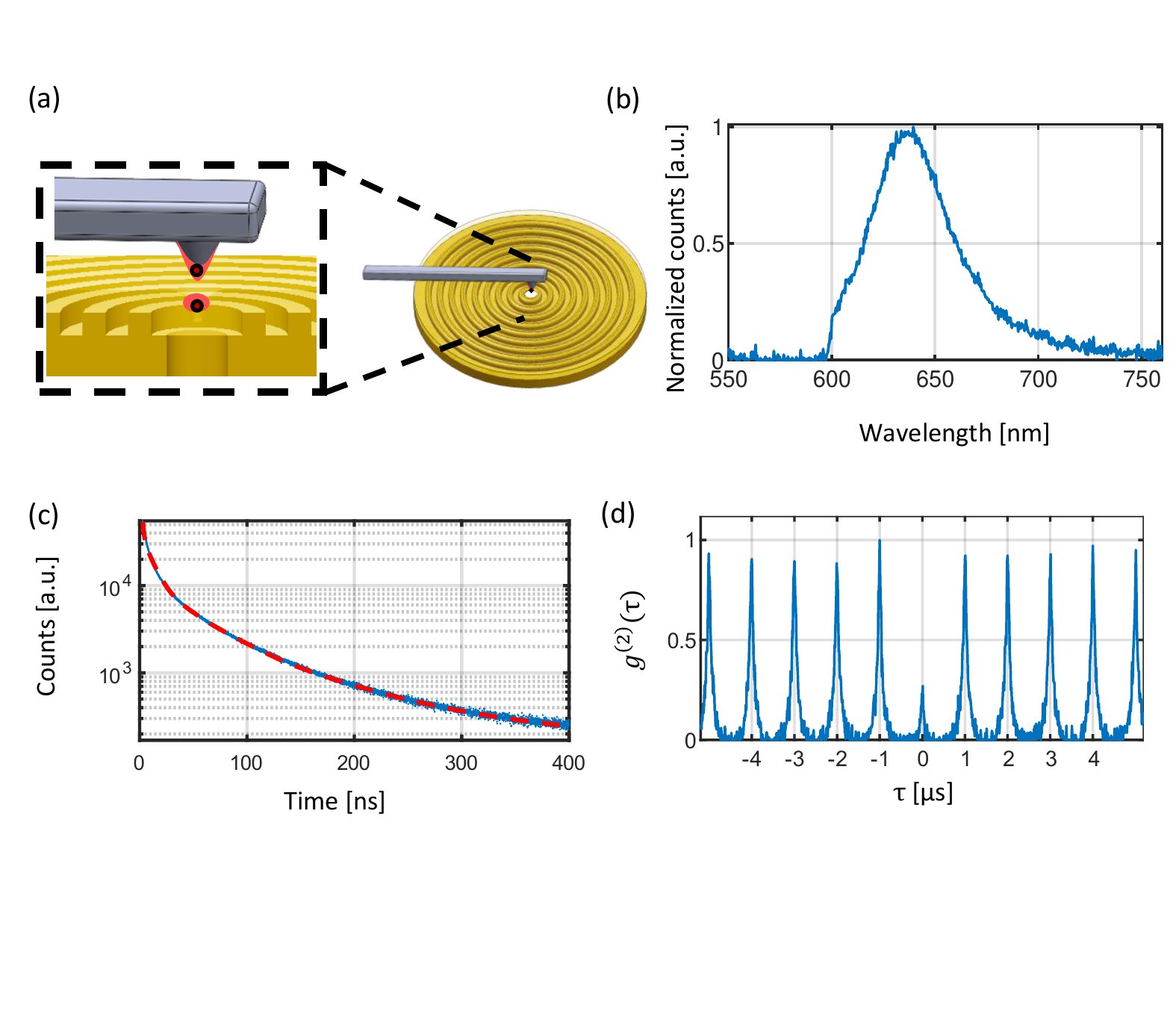}
 \caption{Results from a single QD coupled to bullseye nanoantenna with a hole at the center. (a) Schematic representation of the placement method used for placing the QD on the hole at the center of the bullseye antenna.  (b) QD’s emission spectrum measurement, performed by back excitation. (c) QD fluorescence lifetime measurement on a semi-logarithmic scale, performed by back excitation. (d) Second order correlation measurement using a Hanbury-Brown-Twiss experiment with time-gated filtering technique, displaying an antibunching of less than 0.5. 
}
\label{fig:B}
\end{figure}

Fig. \ref{fig:B} displays the results for a single CQD placed at the center of the nano-antenna. The CQDs used here are CdSe/CdS core/thick-shell nanocrystals, also known as giant CQDs (gQDs). gQDs are characterized by essentially non-blinking and non-photobleaching behavior at room temperature \cite{Chen2008GiantBlinking,Ghosh2012NewDots,Orfield2018PhotophysicsNanocrystals}. This level of stability allows gQDs to provide a reliable quantum source of photons without interruption by fluorescence intermittency or catastrophic failure. The gQDs employed here were synthesized by a modified continuous-injection procedure\cite{Chen2013CompactBlinking} as described in Ref\cite{Orfield2018PhotophysicsNanocrystals} and are characterized by a near-unity  quantum yield of 80\% and monoexponential photoluminescence decay in solution \ref{AppendixB}. 

 A direct-write nanolithography method was used to place individual gQDs into the holes located at the center of the bullseye structures, as shown in Fig. \ref{fig:B}a. 
 This technique is known as dip-pen nanolithography (DPN), as described previously in Ref. \cite{Abudayyeh2021SingleNanoantennas}. DPN is amenable to scaling \cite{DPN} and in previous work demonstrated a 25\% success rate for depositing either a single gQD or a small cluster in standard bullseye antennas.\cite{Abudayyeh2021SingleNanoantennas} In this work, all depositions resulted in placement of either a single gQD or a cluster, as revealed by preliminary second-order correlation ($g^{(2)}(\tau)$) experiments applied using a time-gated filtering technique.\cite{Abudayyeh2021SingleNanoantennas,Mangum2013DisentanglingExperiments} See Appendix B\ref{AppendixB} for details.
 

Fig. \ref{fig:B}b shows the fluorescence spectrum collected by the front objective from a single gQD positioned at the center of the nano-antenna, with a back excitation laser at 405 nm. This typical emission spectrum confirms the successful excitation of a gQD positioned in the front side by the back laser excitation going through the hole. The  lifetime  of the gQDs emission is presented in Fig. \ref{fig:B}c, showing a typical bi-exponential decay characteristic to a bi-exciton - exciton cascade emission from such gQDs on nano-antennas\cite{Abudayyeh2021SingleNanoantennas} 

In order to confirm that the emission is from a single gQD, the results of a time-gated $g^{(2)}(\tau)$ experiment \cite{Abudayyeh2021SingleNanoantennas,Mangum2013DisentanglingExperiments} is shown in Fig.  \ref{fig:B}d.
Here, back optical excitation of a nonresonant ($\lambda$ = 405 nm) pulsed laser at a repetition rate of 1 MHz was used.  With time filtering of $\tau_f=60$ ns a clear antibunching is observed.

\subsection{Fabrication and characterization of a device based on SiV centers in a nanodiamond}

\begin{figure}[H]
\centering
\includegraphics[width=0.7\textwidth]{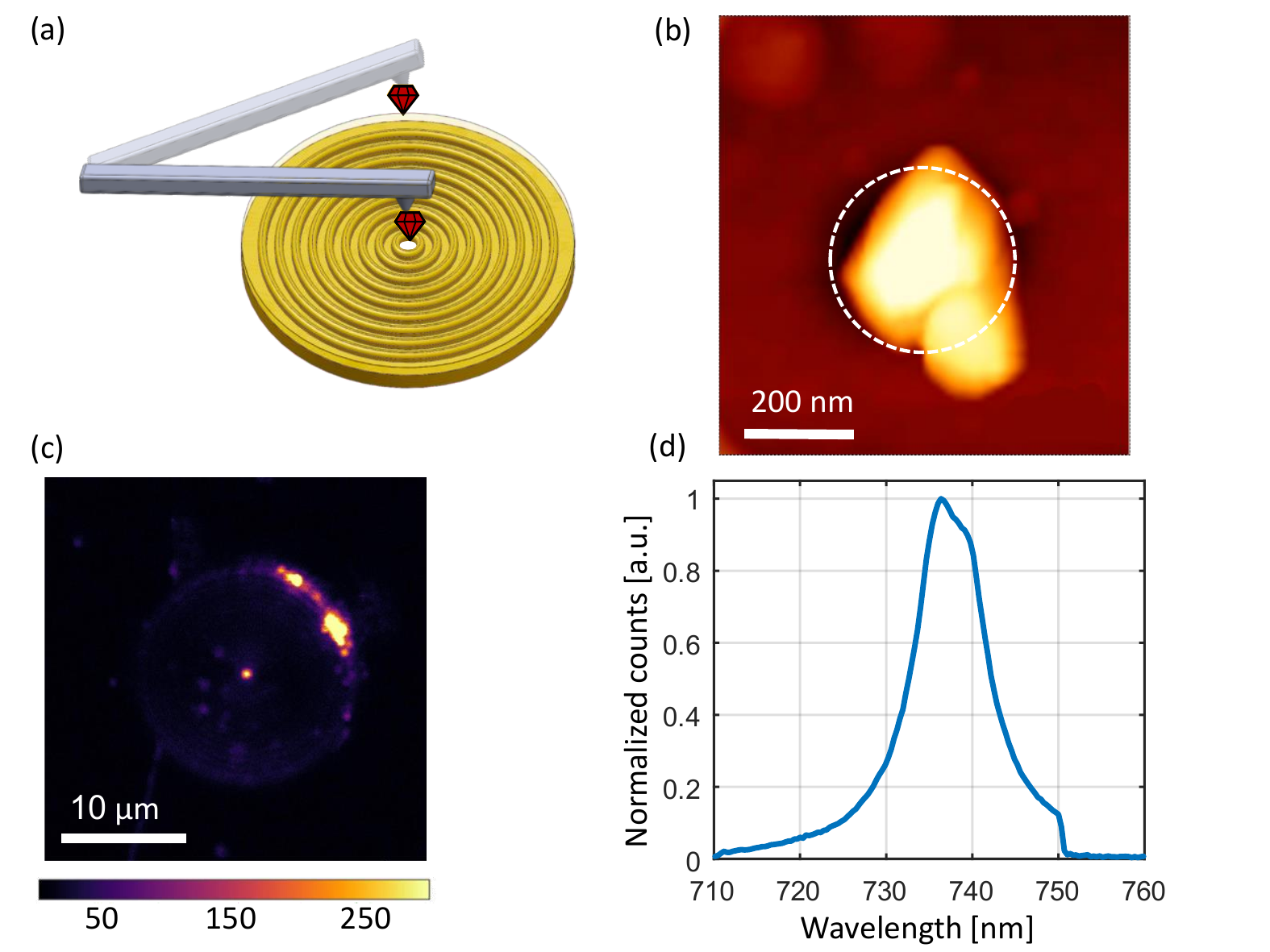}
 \caption{Results from a single ND containing SiV centers coupled to bullseye nanoantenna with a hole at the center.
(a) Schematic representation of the placement method used for placing the ND on the hole at the center of the bullseye antenna. 
(b) AFM scan showing the ND placed on the hole. The white dashed line indicates the hole.
(c) Confocal scan using 532 nm as excitation wavelength of a bullseye antenna showing the emission from the ND located on the hole in the center of the antenna.
(d) SiV center emission spectrum measurement, performed by back excitation.
}
\label{fig:C}
\end{figure}

Symmetry-protected, negatively-charged silicon-vacancy centers (SiV) stand out from the family of color centers in diamond for its dominant photon emission into the zero-phonon line (ZPL) with a Debye-Waller factor of about 70$\%$ as well as spectrally stable and intrinsically indistinguishable optical transitions \cite{Rogers2014a,Hepp2014}. The properties persist even when the SiV center is hosted in crystals at the size of visible wavelengths \cite{Rogers2019SingleCapabilities}. Such nanodiamonds (NDs) offer the possibility of increasing the spin coherence times at cryogenic temperatures by modifying the electron-phonon coupling, thereby suppressing orbital relaxation \cite{Klotz2022ProlongedNanodiamonds}. 
Furthermore, the small size enables precise atomic force microscope (AFM) - based nanomanipulation which opens the way for fabrication of hybrid quantum photonic devices \cite{Fehler2021HybridCavity,Kubanek2022HybridSi3N4-Photonics}.
For Placing the NDs at the center of the bullseye nanoantennas we used the pick and place method which benefits from a high accuracy of  $\sim10$  nm. 
The central hole in the bullseye structure eases the placement of the ND due to its topology.
See Appendix C for details on the preparation and positioning of the NDs.
Fig. \ref{fig:C}b shows an AFM scan of the ND placed on the hole. 
The verification of a successful SiV hosting ND placement is seen in Fig. \ref{fig:C}c.



Fig. \ref{fig:C}d shows the spectrum emitted from a single ND containing SiV centers coupled to the bullseye antenna with a hole at the center. 
The spectrum measurement was performed by back excitation with 532 nm CW laser through the hole, and emission was collected from the front as described in Fig. \ref{fig:A}e. The measurement was taken with a short pass filter (750 nm) which is responsible for the decrease to zero of the intensity at wavelengths above 750 nm.

\subsection{Collection efficiency of CQD and SiV based devices with back excitation}

\begin{figure}[H]
\centering
\includegraphics[width=.7\textwidth]{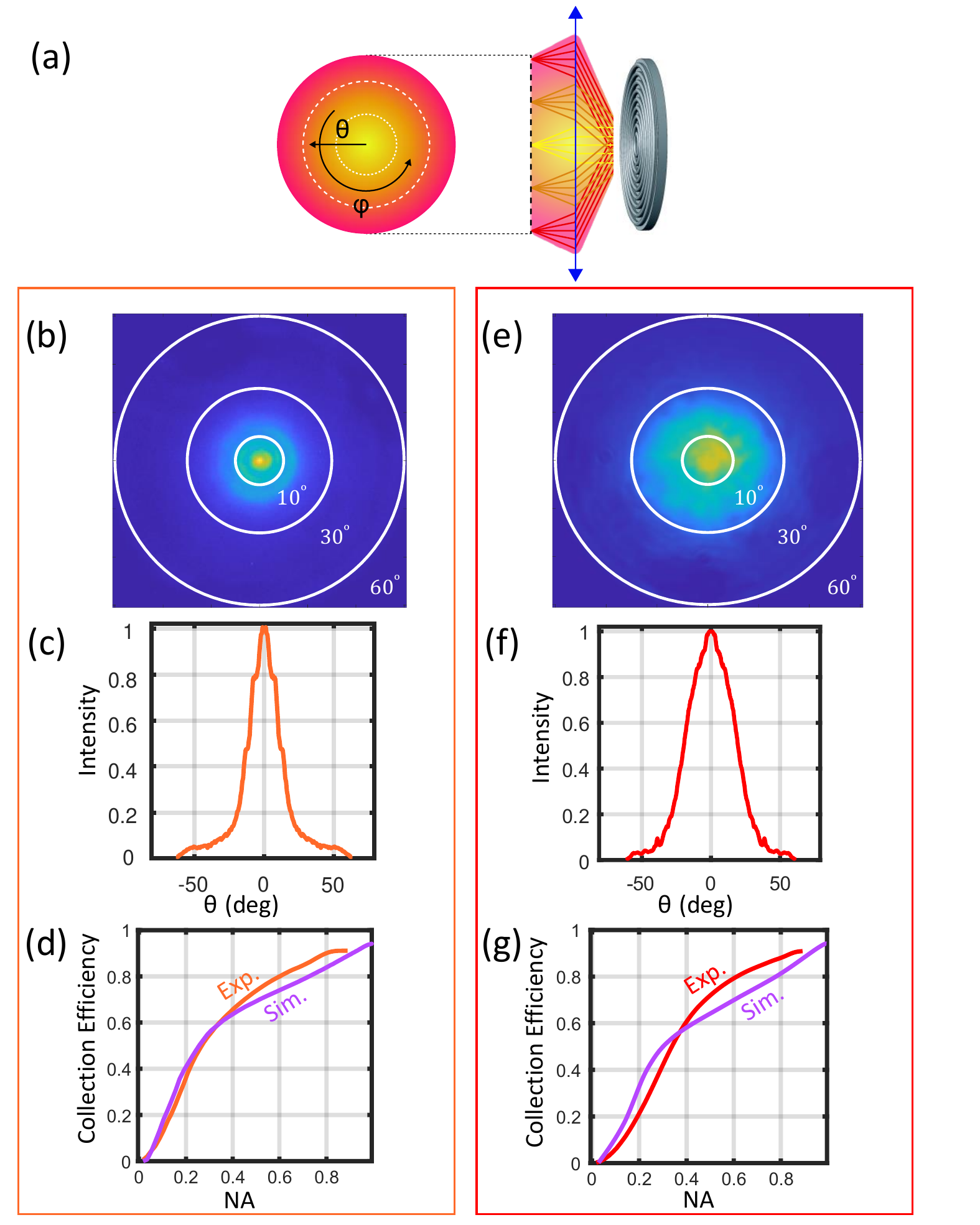}
 \caption{Back focal plane and collection efficiencies measurements and calculations of the device.
(a) Schematic representation of the back focal plane imaging technique used to measure directionality in this study.
The orange frame ((b)-(d)) shows results of the device with single QD and the red frame ((e)- (g)) shows results of the device with single ND containing SiVs.
Back-focal plane image ((b) and (e)), azimuthally integrated angular intensity distribution ((c) and (f)) and simulated (purple) and measured collection efficiency for emitter antenna devices.}
\label{fig:D}
\end{figure}

In order to characterize the performance of both back-excited devices, we measured the PL directionality of the emission. This was done by using Fourier-plane PL imaging, as shown in Fig. \ref{fig:D}a, and results in 2D angular PL intensity distribution map, I( $\theta $ , $\phi $ )
where $\theta $ and $\phi $ are the polar and azimuthal angles, respectively. 
Within the orange frame in Fig. \ref{fig:D}, which includes Fig. \ref{fig:D} b, c, d, we show the results of a single gQD coupled to the antenna.
Fig. \ref{fig:D}b shows the measured back focal plane image which demonstrates highly directional emission. Fig. \ref{fig:D}c presents azimuthally integrated angular intensity distribution, extracted from the back focal plane image. 
The CE $(\eta)$ as a function of NA was extracted by integrating the signal over all azimuthal angles ($\phi $) within a collection cone of a given NA relative to the calculated signal for NA=1, i.e.:
\begin{equation}
 \eta = \frac{\int_0^{2\pi}d\phi\int_0^{\theta_{NA}} d\theta \sin(\theta) I(\theta,\phi)}{\int_0^{2\pi}d\phi\int_0^{\frac{\pi}{2}}d\theta \sin(\theta) I(\theta,\phi)}.
\end{equation}

Fig. \ref{fig:D}d presents the CE as a function of NA for the single gQD based device. This is also compared to FDTD simulation of single linear dipole located at the center of a full device.
Within the red frame in Fig. \ref{fig:D}, which includes Fig. \ref{fig:D}e, f, g, results of single ND containing SiV centers coupled to the antenna are shown. The measured back focal plane image is shown in Fig. \ref{fig:D}e which again demonstrates the emission directionality. 
Fig. \ref{fig:D}f demonstrates the azimuthally integrated angular intensity distribution, extracted from the back focal plane image. Fig. \ref{fig:D}g presents the CE as a function of NA for the SiV-based device. This is again compared to FDTD simulation of single linear dipole located at the center of a full device. We note that for low NAs in the SiV-based device, we measured lower directivity compared to the simulation, as well as compared to the gQD-based device. We attribute the drop in directivity and the lower CE for low NAs to the distribution of the SiV centers within the nanodiamond, which is large compared to the nanodiamond we measured in our previous paper \cite{Waltrich2021High-purityAntenna}.

\subsection{Fiber coupling}

\begin{figure}[H]
\centering
\includegraphics[width=.7\textwidth]{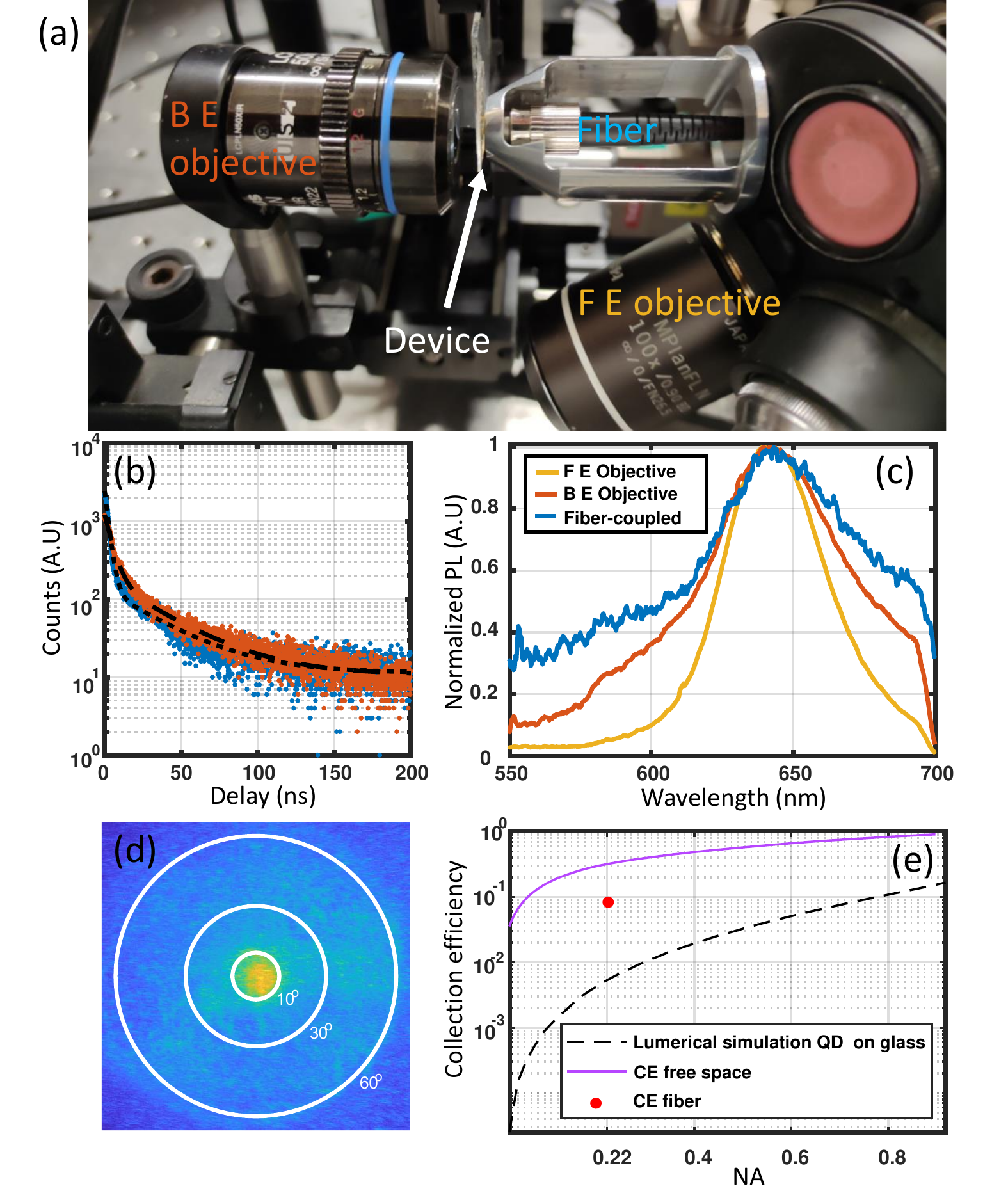}
 \caption{(a) Setup image. (b) Lifetime comparison between an objective collection (orange) and a fiber-coupled (blue) emission. (c) Spectral measurements using three different configurations: front excitation (F E) in free space  (yellow), back excitation (B E) in free space (orange), and back excitation with fiber collection (blue). To isolate specific spectral regions, spectral filters were placed at 550 and 700 nm (d) Back focal plane image of the same device. (e) Comparison of back excitation collection efficiency (magenta) and QD on glass collection efficiency (black). 
 }
\label{fig:E}
\end{figure}

Together with the very high directionality of the front emission of photons, one of the main advantages of the back excitation scheme together with the very high directionality of the front emission of photons, is the prospect of an integrated photonic device where the single-photons emitted are directly coupled into a proximal optical fiber without the need for complex, bulk optical elements in between. Here we demonstrate the ability to perform such direct coupling. And, we do so at room-temperature in contrast with prior studies. Fig. \ref{fig:E} shows the results of the fiber coupling measurements for a gQD device. For simplicity, the experiment was performed with a back-excited device that contained several gQDs at the center of the nano-antenna(Appendix B \ref{AppendixB} Fig. \ref{fig:appendix b}h).
In Fig \ref{fig:E}a we present a photo of the experimental setup for direct coupling, having the same back excitation objective as in the free space measurements. At first, the fiber position with respect to the nano-antenna was aligned using the piezo-electric stage to optimize the direct coupling to the fiber (see Fig \ref{fig:E}a). Afterward, two measurements were conducted, both directly through the fiber and with a front objective: (1) a PL lifetime measurement and (2) a PL spectrum measurement. In Fig.  \ref{fig:E}b we present a comparison between two lifetime traces, one (orange) was measured using the 0.9 NA objective and the other (blue) was measured through a 0.22 NA MM fiber. Both lifetime traces show the typical bi-exponential decay of our gQDs. Fig \ref{fig:E}c shows a comparison of three normalized spectroscopic measurements: back excitation and front collection using a 0.9 NA objective, front excitation and front collection with a 0.9 NA objective, and back excitation with a direct 0.22 NA fiber collection. As seen in the figure, all three measurements display a similar spectrum typical to the gQD, having a central peak near 645 nm. These results prove that indeed the gQDs emission was successfully coupled directly into the fiber. Fig \ref{fig:E}d shows the PL angular intensity map of the device under back-excitation. A directionality into an angular spread approximately corresponding to the NA of our MM fiber is observed, which we argue is the reason for successful direct fiber coupling. To estimate the coupling efficiency into free space and directly into the fiber, we conducted a photon count comparison, taking into account different losses of the collection optical setup (see Appendix D for details). This is presented in Fig. \ref{fig:E}e. A coupling efficiency of $9.88\pm 0.48\% $ was found. We attribute the difference between this value to the CE at the same NA in free space to imperfect alignment, mode mismatch between the free space and the MM fiber, and dielectric reflection from the uncoated fiber facet. This preliminary result opens a path towards on-chip single-photon devices with a direct fiber coupling integrated with an excitation source from the back.
\subsection{Discussion and conclusion}
We demonstrated the emission of single-photons from a back excited on-chip emitter - nano-antenna device, displaying efficient back-excitation, high forward directivity and high collection efficiency.  Devices of SiV-centers in NDs and gQDs, placed on a metal-dielectric bullseye antenna with a subwavelength hole at the center were demonstrated.
Remarkably, the drop in CE compared to similar devices without a hole and with front excitation \cite{Abudayyeh2021SingleNanoantennas,Abudayyeh2021OvercomingSourceb} as a result of adding the hole amounts to less than 10$\%$ for an NA of 0.5. We attribute this minor drop in CE to emission losses into the hole, as well as less effective emission coupling to the antenna's spatial modes. Further optimization of the device, by reducing the diameter of the hole, as well as increasing the depth of the hole by evaporating a thicker Au layer should further increase the collection efficiency.

Such on-chip RT single-photon devices with high CE are very appealing for compact integration into free space and fiber based quantum communication systems\cite{Schimpf2021QuantumDots,Bayer2021SemiconductorChallenges, KimSemiconductor1,BozzioEnhancingSources,Sipahigil2016AnNetworks,Nguyen2019AnDiamond,Knall2022EfficientSystem}.
We also demonstrated the applicability of highly accurate positioning methods of nano-emitters for complex device fabrication: for NDs containing SiV-centers, we utilized a pick and place method, which is highly accurate and enables multi-stage position optimization and dipole rotation optimization \cite{Hauler2019PreparingFreedom}. 
For the QDs placing, we utilized a DPN technique, which is amenable to scaling and is highly accurate\cite{Abudayyeh2021SingleNanoantennas}. 
The compatibility of the nano-antenna device concept for two different quantum light sources using two different coupling methods emphasizes the versatility and robustness of this system, and indicates the variety and range of possibilities of quantum technologies to which this device can be adapted. The successful direct coupling to a low NA fiber is promising as a platform for future fiber coupled on-chip RT SPS systems. 

\section{ Acknowledgments}
The authors thank V.A. Davydov for synthesis and processing of the nanodiamond material. The authors also thank Patrick Maier for preparing the FIB markers. The gQD synthesis, integration and preliminary quantum-optical characterization were performed at the Center for Integrated Nanotechnologies (CINT), a Nanoscale Science Research Center and User Facility operated for the U.S. Department of Energy (DOE) Office of Science. The fabrication and characterization of the nano-antenna devices were performed at the Hebrew University center for Nanoscience and Nanotechnology. The project was partially funded in a Quantum Alliance project within IQst.
The project was partially funded by the BMBF/VDI within the project HybridQToken (16KISQ043K). A.K. acknowledge support of IQst. R.R. acknowledges support from the Quantum Communication consortium of the Israeli Innovation Authority. A.V.B. and E.G.B. were funded by CINT. All other Los Alamos National Laboratory (LANL) authors were funded by the Laboratory Directed Research and Development (LDRD) program (Grant No. 20170001DR), and J.A.H. acknowledges funding through the U.S. DOE, Office of Science, Office of Advanced Scientific Computing Research, Quantum Internet to Accelerate Scientific Discovery Program.

\bibliography{main.bib}
\newpage
\input{AppendixA}

\newpage
\input{AppendixB}

\newpage
\input{AppendixC}

\input{AppendixDV2.tex}

\end{document}

%% file: AppendixA.tex
\section{Appendix A - Design and fabrication of the nano-antenna devices}
We used the template stripping method in order to fabricate and achieve high quality bullseye antennas, as reported previously \cite{Abudayyeh2021OvercomingSource,Abudayyeh2021SingleNanoantennas}.
With this method, after the stripping we will be left with a device that consists of a sapphire slide on top of which are two transparent layers -
SU8 3010 photo-resist with a thickness of $\sim$ 10 microns, and on top a layer of gold with a 
thickness of 250 nm.
We used a Focused Ion Beam (FIB) machine to drill a hole in the center of the antenna. 
The FIB allowed us to drill a hole in a controlled manner so that it was possible to detect when the gold layer was removed and immediately stop. In this way we could remove the gold layer and be left with only the two bottom layers which are transparent.
For a hole with a diameter of 400 nm we used a  30 kV ion beam with 7.7 pA current. 

In order to determine the desired size of the hole, we used a Lumerical simulation. 
As mentioned in the main text, the hole size must be large enough to allow transmission of the laser light through the hole and excite
 the dipole, and small enough to maintain the CE by reducing the transmission of the light
 emitted from the dipole source. 
For this purpose, we performed a simulation of the laser transition and the CE as a function of the hole diameter, and thus we could choose a size that would match both requirements, as shown in the in Fig. \ref{fig:A}d in the main text.
The simulation was based on the model described in  \cite{Abudayyeh2017} with the addition of the hole with different diameter values.
Before placing the emitter on the antenna the sample was covered with a dielectric layer of aluminum oxide using ALD deposition which allows us to determine with nanometric precision the thickness of the dielectric layer.
We measured the thickness of the aluminum oxide layer in the region of the hole and the hole size ($d$) using AFM as illustrated in the inset of Fig. \ref{fig:A}a. 
The optimized values of our samples, which are shown in Fig. \ref{fig:A}a are $d$ =400 (400)nm, $D$=1260 (1430)nm, $\Lambda$=560 (635)nm, $h$= 215 (243)nm for the CQD (SiV) respectively.

%% file: AppendixB.tex
\section{Appendix B - Characterization of gQDs and Positioning by Dip-pen Nanolithography}\label{AppendixB}

\begin{figure}
    \centering
    \includegraphics[width = 0.8\textwidth]{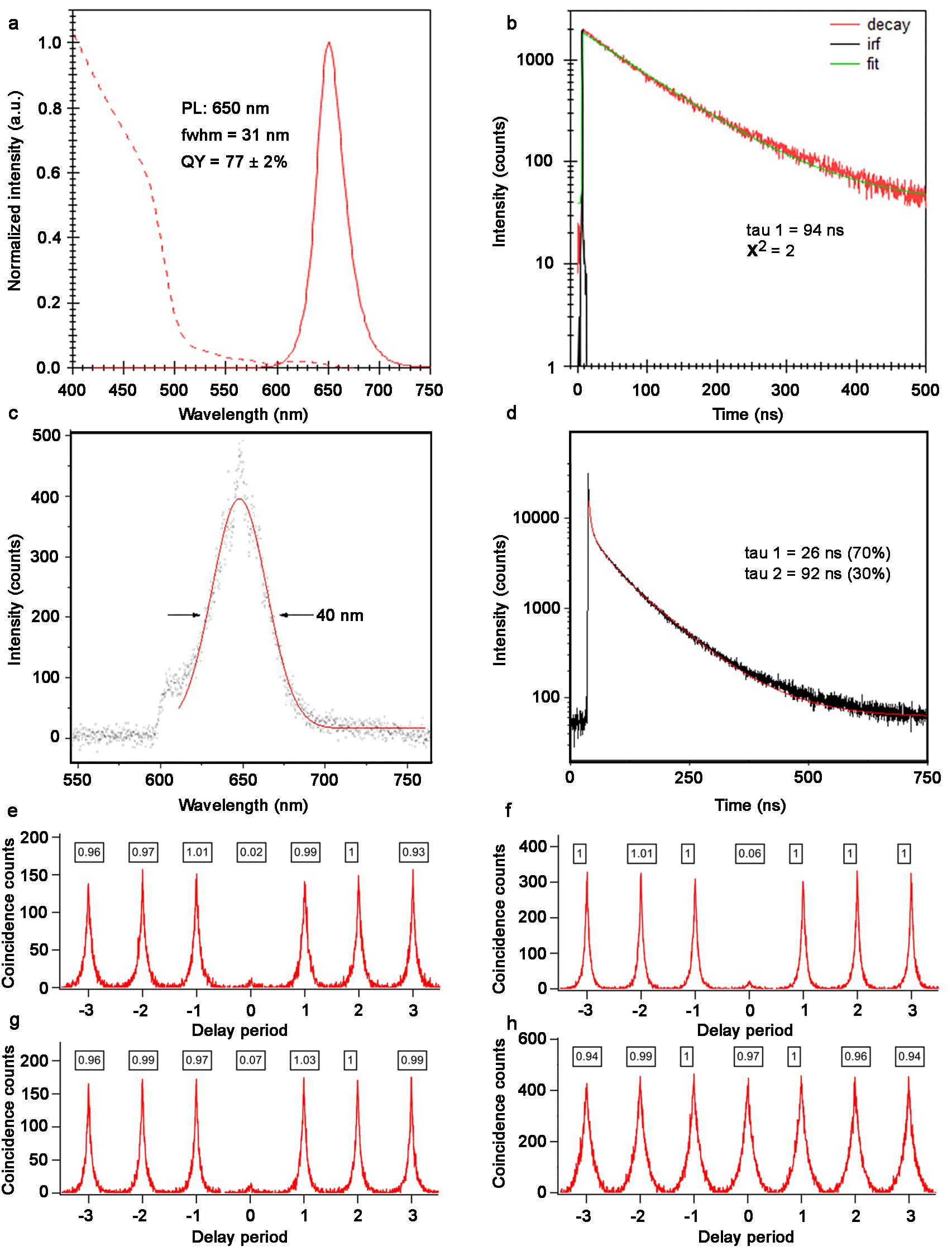}
    \caption{(a) Absorption and emission spectra and (b) PL decay lifetime obtained from gQDs suspended in hexanes. (c) PL spectrum for a single gQD embedded in a bullseye antenna. (d) PL decay lifetime for the single gQD in (c). (e) second-order fluorescence intensity correlation ($g^{(2)}$) for the single gQD in (c). (f) and (g) Correlation data for two other single-gQD/antenna couples. (h) Correlation data for a gQD cluster/antenna couple. In (e)-(h) a time gating of 50 ns was employed to remove contributions from faster biexciton emission (compared to slower exciton emission). Correlation data were used after deposition by dip-pen nanolithography to assess whether single or multiple gQDs were placed. Since $g^{(2)}(0)$ is still large after time-gating for the gQD-antenna device in (h), multiple gQDs are understood to be present. }
    \label{fig:appendix b}
\end{figure}

Prior to integration into bullseye antennas, gQDs were characterized in the solution phase for absorption and photoluminescence (PL) properties (Figure \ref{fig:appendix b}a, b). Steady-state emission spectra, lifetime, and quantum yield (QY) measurements were made using an Edinburgh Instruments FLS1000 fluorescence spectrometer equipped with 450 W xenon lamp (steady-state emission spectra and QY measurements) and AGILE supercontinuum laser (lifetime measurements) excitation sources. Absorption measurements were made with a Cary 5000 UV-Vis-NIR spectrometer. PL QY was determined using a standard Edinburgh Instruments integrating sphere. QY measurements on hexane dispersions of CdSe/CdS gQDs in quartz cuvettes were made in triplicate, and absorbance spectra were acquired on all samples to confirm that the absorbance value at the excitation wavelength was $<$ 0.1 to limit self-absorption effects. The sample was excited at a wavelength of 490 nm with a monochromator grating of 6.00 nm, and data were collected with an emission monochromator grating of 0.1 nm. The resulting QYs obtained over three measurements afforded a value of 77 ± 2\%. PL lifetime measurements were made at an excitation wavelength of 400 nm with a pulse repetition rate of 1 MHz, and the detection wavelength was set to 650 nm (center of the gQD emission signal). Excitation and emission monochromator gratings were set to 3.00 nm and 1.00 nm, respectively. The resulting PL decay curve is shown in Fig. \ref{fig:appendix b}b.

The technique used to place individual gQDs into the holes located at the center of the bullseye structures is known as dip-pen nanolithography (DPN). It allowed us to transfer gQDs that are suspended  in a high-boiling solvent (dichlorobenzene) and wicked onto the tip of an atomic force microscopy (AFM) cantilever from the AFM tip to the target substrate. The general method was described previously in Ref. \cite{Abudayyeh2021SingleNanoantennas}, and modified slightly here as follows. The AFM tips were placed in the gQD suspension for $\sim$30 s, and excess ink was removed prior to writing by conducting a series of quick scans on a region of the substrate well away from the bullseye structures until ink could no longer be seen coming off of the tip in the DPN’s 10x optical microscope. The employed tip-surface contact times for writing into the hole containing bullseyes was 0.20 - 0.25 s. While it is certain that the writing tip successfully targeted the bullseye center in each placement attempt, it was not possible to determine whether the tip made contact within the hole or at the top surface or hole edge. Nevertheless, in each of four attempts gQDs were successfully delivered into a hole, with 3 receiving single - gQDs and 1 a small cluster of gQDs. By stating this, we do not mean to imply that DPN can deliver small numbers of nanocrystals with 100\% certainty, as the attempted depositions are too few from which to draw such a conclusion. In our previous work studying hole-free bullseyes, we demonstrated a 25\% success rate for depositing either a single gQD or a small cluster,\cite{Abudayyeh2021SingleNanoantennas} and it remains to be determined whether the hole itself plays a role in directing fluid movement on the surface of the device, essentially assisting in the placement process.

After placement into the antennas, we immediately characterized the deposited gQDs using a combination of single-emitter PL spectroscopy, PL lifetime evaluation, and $g^{(2)}(0)$ determination (Fig. \ref{fig:appendix b} c-h). This allowed us to confirm the success of our deposition, including whether single or multiple gQDs were deposited. Specifically. a picosecond pulsed laser (Picoquant LDH-D-C-405) with a wavelength of 405 nm and pulse width of 56 ps was used for excitation. The laser was reflected through a dichroic beamsplitter (Semrock Di02-R405) and then focused onto the sample to a diffraction limited spot size using a 50×, 0.7 NA Olympus objective microscope (LCPLFLN50xLCD), which was used to both excite the sample and collect the PL. Collected photons after passing through the same dichroic beamsplitter and a 590 nm long-pass filter either go to a spectrometer + CCD (Acton SP2300i, pylon100) or a Hanbury Brown-Twiss setup consisting of a 50:50 beamsplitter and two single photon avalanche photodiodes (Excelitas SPCM-AQRH-14). TRPL was analyzed using a TCSPC module (Picoquant Hydraharp 400). Excitation power before the objective is about 500 nW. 

%% file: AppendixC.tex
\section{Appendix C - Preparation and positioning of Nanodiamonds using Pick-and-Place}
In preparation of transferring SiV hosting NDs to the bullseye antenna, a dispersion of SiV containing NDs was drop-casted onto a fused silica coverslip.
The NDs were fabricated using high-pressure high temperature (HPHT) treatment of the catalyst metals-free hydrocarbon growth system. The NDs were
treated with \ch{HNO3 + HClO4 + H2SO4} and HF to remove sp2 carbon and were
washed and dried afterwards. FIB milled marker structures on the fused silica coverslip serve as a position reference to later locate specific SiV-containing NDs
with an AFM. To find SiV-hosting NDs the sample is examined using a homebuilt confocal microscope, orchestrated based on the open-source software qudi \cite{Binder2017Qudi:Processing}. A 2D galvo scanning mirror, a 4f-system and a 1.35 NA oil objective form the basis of this optical setup. Narrowband filtering (740 ± 13 nm)
around the ZPL of the SiV increases the signal to noise ratio when exciting offresonantly with 532 nm, as the dominant emission originates from SiV centers
within this filter window. After suitable NDs have been located, AFM imaging
of the area of interest is carried out. Triangulation of the ND position is done
with the help of the FIB markers, as well as positions of other fluorescing NDs
in the confocal image. In order to pick and place \cite{Schell2011ADevices} the ND of interest, a platin-coated AFM cantilever is approached with a constant force until the ND attaches to the tip. The pick-up is indicated by either the disappearing ND in a subsequent non-contact AFM scan or by image artifacts such as ’double-tip’ features. Those features also hint towards a successful placement strategy of the ND in the following step. With the ND attached to the cantilever tip, the sample is exchanged for the target sample, here the bullseye antenna. A careful non-contact imaging reveals the target position on the structure. For the antenna structure considered herein the central hole in the structure eases the placement of the ND due to its topology. A contrast in height is useful to detach the ND from
the cantilever. After successful ND placement (compare Fig. \ref{fig:C}b), the sample is again examined in the confocal microscope, with the objective being exchanged to 0.55 NA for the sake of a longer working distance and to leverage the collection efficiency \cite{Waltrich2021High-purityAntenna}. The verification of a successful SiV hosting ND placement is seen in Fig. \ref{fig:C}c.

\label{AppendixC}

%% file: AppendixDV2.tex
\section{Appendix D - Estimation of in-fiber collection efficiency}

\begin{figure} [H] 
    \centering 
    \includegraphics[width = 0.7\textwidth]{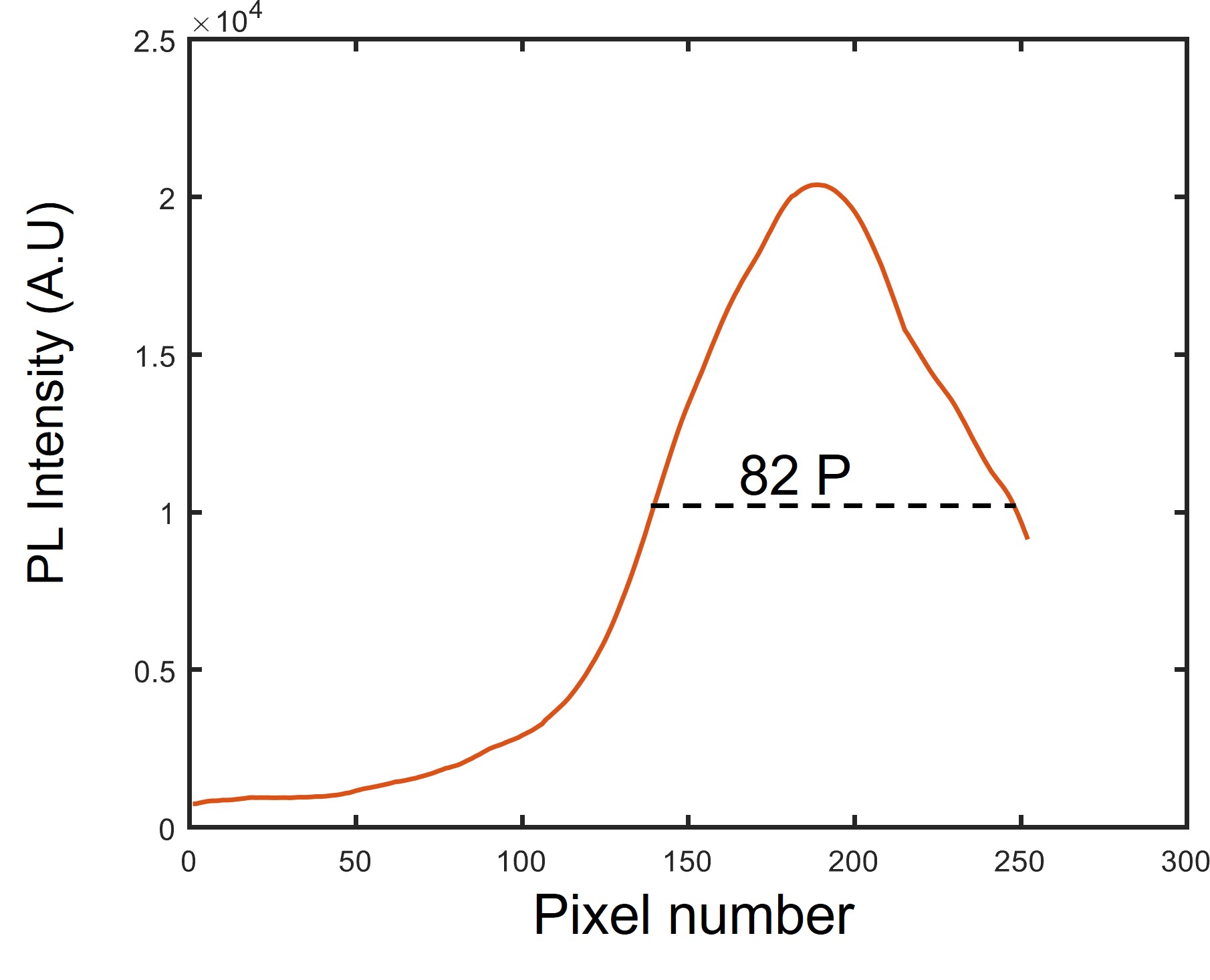} 
    \caption{FWHM of the intensity distribution on the spectrometer camera}
    \label{fig:pixel} 
\end{figure}

To estimate the in-fiber collection efficiency, we calculated the fraction of collected photons from all emitted photons using back excitation. First, using the back focal plane image(\ref{fig:E}.d), the relative collection efficiency into NA = 0.22 (which is the NA of our MM fiber) was calculated to set the upper limit of the achievable coupling using this device. 
\begin{equation}
   \frac{\eta_{NA = 0.22}}{\eta_{NA = 0.9}} = \frac{ \int_{0}^{\theta_{NA = 0.22}}d\theta sin(\theta) \int_{0}^{2\pi} I(\theta,\phi)d\phi}{\int_{0}^{\theta_{NA = 0.9}}d\theta sin(\theta) \int_{0}^{2\pi} I(\theta,\phi) d\phi}
\end{equation}
Where $\eta$ is the collection efficiency and $I(\theta,\phi)$ is the intensity in the back focal planee image. we find
\begin{equation}
   \eta_{NA = 0.22} = 0.32   \qquad  \eta_{NA = 0.9} = 0.91
\end{equation}
and thus,
\begin{equation}
    \frac{\eta_{NA = 0.22}}{\eta_{NA = 0.9}} = 0.3516 
\end{equation}
Hence, $35.19 \% $ of the light collected by our 0.9 NA objective is emitted at angles smaller than $\theta_{NA = 0.22}$. Next, to estimate the experimental coupling efficiency, we compared the counts obtained on a Pixis spectrometer camera for the whole emission spectral band using a collection with a 0.22 NA fiber and a 0.9 NA objective.
\begin{equation}
    \frac{I_{Fiber\: Coupled}}{I_{0.9 NA\: objective}} = \frac{\int_{\lambda = 550 nm}^{\lambda = 700 nm} I_{Fiber\: Coupled}(\lambda) d\lambda \cdot \gamma_{Fiber}^{-1}}{\int_{\lambda = 550 nm}^{\lambda = 700 nm} I_{0.9 NA\: objective}(\lambda) d\lambda \cdot \gamma_{Objective}^{-1}}
    \label{equation:fiberc}
\end{equation}
where the integral represents the area under the spectrum and $\gamma$ stands for the total measured loss in each collection setup (free space and fiber collection setups). The losses in the objective system are divided into two distinct types:
\begin{equation}
    \gamma_{objective} = \gamma_{system \: loses} \cdot \gamma_{spectrometer \: slit}
\end{equation}
where the losses in the system,$\gamma_{system \: loses}$, are losses caused by the optical system itself (partial transmission of mirrors and filters) and losses caused by the spectrometer's entrance slit,$\gamma_{spectrometer \: slit}$. Since the entrance slit is slightly smaller than the entire emission spatial beam, only a portion of the light that reaches the rectangular slit is transmitted. To calculate the fraction of transmitted light, we use the following expression:
\begin{equation}
    \gamma_{spectrometer \: slit} = \frac{\Gamma w}{\pi \cdot \Gamma^2}
    \label{equation:slit}
\end{equation}
Where $\Gamma$ is the FWHM of the Gaussian emission profile on the spectrometer camera along the slit axis, and $w$ is the slit width. To calculate the FWHM of the power distribution, we sum the intensities along the camera and not along the grating. As seen in the figure below \ref{fig:pixel} the FWHM is 44 pixels.
Since the pixel size is 82 microns, the FWHM is $2132 \unit{\mu m}$. By using the known width of the spectrometer slit which is $500 \unit{\mu m}$, we can substitute it into \ref{equation:slit}:
\begin{equation}
    \gamma_{spectrometer \: slit} = \frac{2132 \unit{\mu m} \cdot 500 \unit{\mu m}}{\pi \cdot (2132 \unit{\mu m})^2} \approx 0.074
\end{equation}

To determine the loss coefficient for the objective system, we used a He:Ne laser. By measuring the power before the objective and after all the mirrors and filters, we found that the measured power was $6.65 \unit{\mu W}$ and $17 \unit{\mu W}$, respectively. Therefore the loss coefficient for the objective system is approximately $\gamma_{system : loses} \approx 0.3911$. 

\begin{figure} [H] 
    \centering 
    \includegraphics[width = 0.7\textwidth]{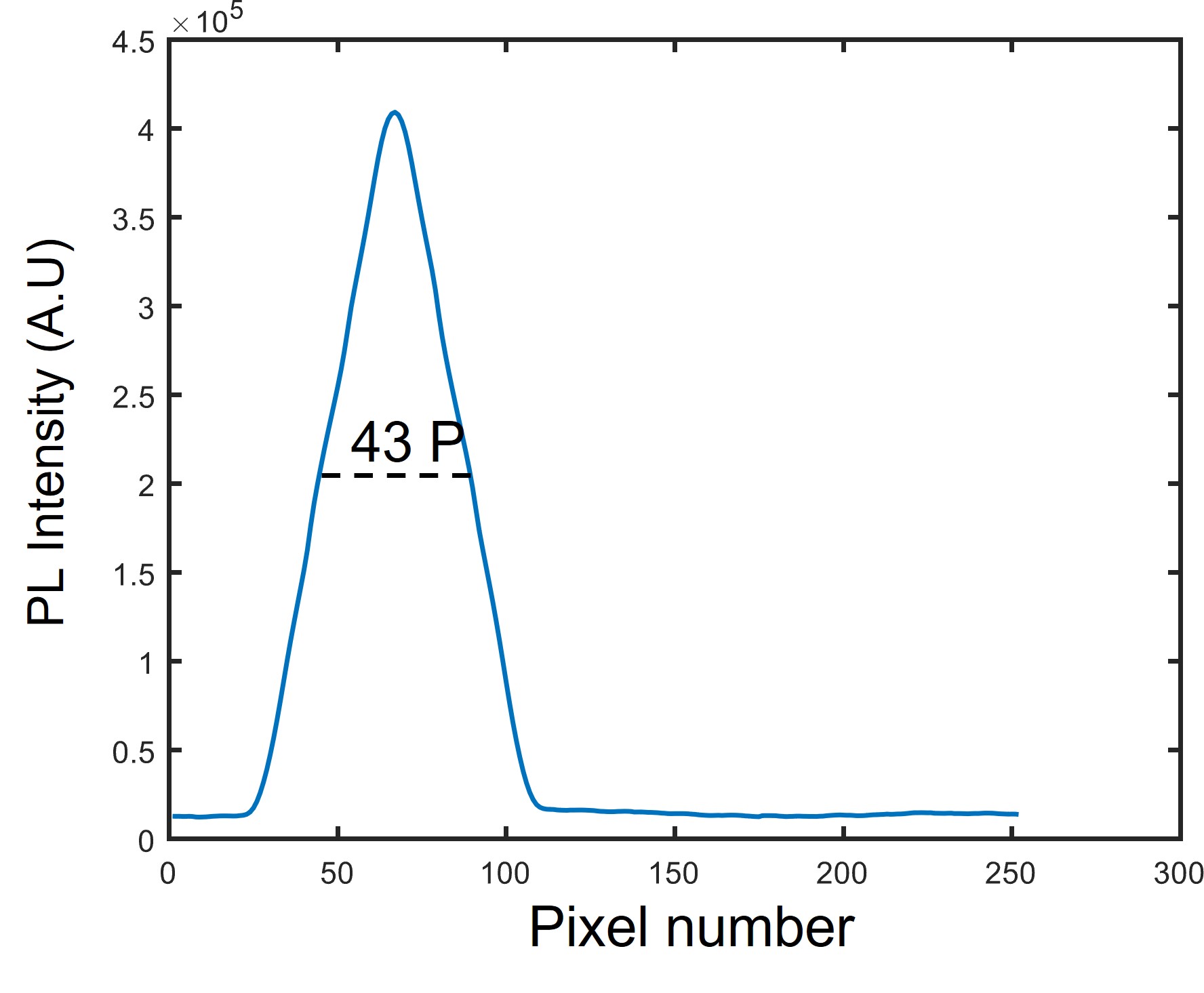} 
    \caption{fiber measurement: FWHM of the intensity distribution on the spectrometer camera}
    \label{fig:pixel2} 
\end{figure}

The losses in the fiber coupling system were calculated using the same He:Ne laser. The intensity of the laser was measured before the entrance to the fiber. After the filters on the other side of the fiber, the intensity before the entrance was $17 \unit{\mu W}$ and $14.6 \unit{\mu W}$ at the end of the system (after all of the mirrors and filters).
Hence, the loss of the fiber system was  $\gamma_{fiber \: loss} \approx 0.859$.
Here again, we must find the loss due to the spectrometer slit.
\begin{equation}
 \gamma_{Fiber} = \gamma_{fiber \: system} \cdot \gamma_{fiber \: spectrometer\:slit} 
\end{equation}
Again, we will find the FWHM of the intensity distribution on the camera. 
As it can be seen, the FWHM is 43 pixels\ref{fig:pixel2}; multiplying that by the size of the pixels, we get 1118 microns.
 
\begin{equation}
        \gamma_{spectrometer \: slit} = \frac{\Gamma w}{\pi \cdot \Gamma^2}
        = \frac{500 \cdot 1118}{\pi \cdot (1118)^2} = 0.1423
    \label{equation:slit2}
\end{equation}
and hence,
\begin{equation}
    \gamma_{Fiber} = 0.0854 \cdot 0.859 \approx 0.1222
\end{equation}
Now substituting all the values into the previous equation \ref{equation:fiberc}. 
\begin{equation}
    \frac{I_{Fiber\: Coupled}}{I_{0.9 NA\: objective}} \approx \frac{\int_{\lambda = 550 nm}^{\lambda = 700 nm} I_{Fiber\: Coupled}(\lambda) d\lambda \cdot 8.177}{\int_{\lambda = 550 nm}^{\lambda = 700 nm} I_{0.9 NA\: objective}(\lambda) d\lambda \cdot 34.25} = \frac{98300 \cdot 8.177}{237500 \cdot 34.25}  \approx 0.0988
    \label{equation:fiberc2}
\end{equation}
And hence the effective coupling efficiency is  $9.88 \pm 0.48 \%$.
\label{AppendixA}